# Direct Measurement of Zak Phase and Higher Winding Numbers in an Electroacoustic Cavity System

Guang-Chen He[1], Zhao-Xian Chen[2,3,4], Xiao-Meng Zhang[1], Ze-Guo Chen[1,3,*], Ming-Hui Lu[2,3,4]

[1]School of Materials Science and Intelligent Engineering, Nanjing University, Suzhou 215163, China

[2]College of Engineering and Applied Sciences, Nanjing University, Nanjing 210093, China

[3]National Laboratory of Solid State Microstructures, Nanjing University, Nanjing 210093, China

[4]Collaborative Innovation Center of Advanced Microstructures, Nanjing University, Nanjing 210093, China

*Corresponding author.

E-mail addresses: zeguoc@nju.edu.cn (Ze-Guo Chen).

## Abstract

Topological phases are states of matter defined by global topological invariants that remain invariant under adiabatic parameter variations, provided no topological phase transition occurs. This endows them with intrinsic robustness against local perturbations. Experimentally, these phases are often identified indirectly by observing robust boundary states, protected by the bulk-boundary correspondence. Here, we propose an experimental method for the direct measurement of topological invariants via adiabatic state evolution in electroacoustic coupled resonators, where time-dependent cavity modes effectively emulate the bulk wavefunction of a periodic system. Under varying external driving fields, specially prepared initial states evolve along distinct parameter-space paths. By tracking the relative phase differences among states along these trajectories, we successfully observe the quantized Zak phase in both the conventional Su–Schrieffer–Heeger (SSH) model and its extended variant with next-nearest-neighbor coupling. This approach provides compelling experimental evidence for the precise identification of topological invariants and can be extended to more complex topological systems.

**Introduction**

Since the emergence of concepts like topological insulators and the quantum spin Hall effect, investigations into topological phases have become a central research topic in condensed matter physics and quantum information science[1,2]. In contrast to traditional phase transitions reliant on symmetry breaking and local order parameters, topological phases are defined by the intrinsic topological structure of the system's parameter space[3,4]. To characterize and differentiate topological phases, researchers have introduced a range of topological invariants, including the well-known Chern number[5] and the Berry phase[6]. In one-dimensional periodic systems, the Berry phase, referred to as the Zak phase[7], dictates whether the system can support topologically protected boundary-localized states under open boundary conditions. This phenomenon illustrates the bulk-edge correspondence in topological systems[8,9], demonstrating a precise relationship between the topological invariants of bulk states and the boundary states that emerge at the edges. In most cases, experimental studies infer bulk topological properties indirectly through boundary state measurements in diverse platforms, including cold-atom systems (by flexibly tuning the quasi-lattice potential)[10,11], photonic crystals (by adjusting the lattice spacing)[12,13], acoustic systems (by modifying periodic structural parameters)[14,15], and superconducting circuits (by configuring microwave resonator networks)[16,17].

Observing boundary states merely provides an indirect approach to inferring a system's topological properties. For a more direct measurement, one must ensure that a wave function with a well-defined momentum undergoes adiabatic evolution across the entire Brillouin zone, enabling the extraction of topological invariants such as the Berry phase. Over the past decade, several groups have attempted to realize such direct measurements in various platforms—ranging from ultracold atoms in optical lattices[18,19] to mechanical oscillators[20], electromechanical cavity systems[21]—by engineering controlled adiabatic cycles in momentum space. However, achieving this momentum-space traversal imposes strict requirements on phase-coherent control of the wave function in real space. Moreover, in practical systems, residual symmetry breaking[7,22], disorder[23-25], or systematic errors[26] can cause the topological phase to lose its quantized character, making the direct measurement of topological invariants even more challenging.

In this work, we propose a method for directly measuring the Zak phase by constructing an equivalent Hamiltonian and controlling the time evolution of its eigenstates. We design a time-varying system[27,28] that maps the wave function from the target system's momentum space onto a high-dimensional real parameter space. Then, we induce an adiabatic evolution of the eigenstates along a chosen trajectory in this parameter space. Through extracting the geometric phase accumulated during wave function evolution along various prescribed trajectories in the Brillouin zone, we successfully measured the Zak phase in both the standard SSH model and an extended SSH model with next-nearest-neighbor coupling. Notably, we observed a phase-doubling phenomenon in the extended model. The key advantage of our approach lies in its ability to precisely control wave function evolution in a highly regulated environment while reducing reliance on external reference gauges, thereby providing a direct experimental paradigm for measuring topological phases. Furthermore, this framework can be extended to higher-dimensional, nonlinear, or long-range coupling scenarios, offering a robust experimental platform for exploring more complex topological systems.

**Results**

**Zak Phase Measurement of the SSH Model**

Consider the SSH model depicted in Fig. 1(a), whose momentum-space is given by[29] (see Fig. 1(b))

$$H(k) = \begin{bmatrix} 0 & w + ve^{-ik} \\ w + ve^{ik} & 0 \end{bmatrix}, \quad (1)$$

where $w$ and $v$ are the coupling coefficients that alternate between adjacent lattice sites. Within this framework, an important characterization of topological properties is the Zak phase[7], which is defined in the momentum-space as

$$\phi_g = \int_{BZ} i\langle u_n(k)|\partial_k u_n(k)\rangle\, dk, \tag{2}$$

where $|u_n(k)\rangle$ denotes the normalized eigenstate corresponding to the $n$th eigenvalue level. $\phi_g$ denotes the geometric phase accumulated by the wave function as the momentum $k$ traverses the entire Brillouin zone. To achieve effective experimental control over the Hamiltonian originally dependent on momentum $k$, we treat $k$ as a time-evolving parameter, defined by $k(t) = (-\pi + \frac{2\pi}{T}t)(t \in [0,T])$. In this way, the $k$-dependent Hamiltonian is transformed into a time-dependent Hamiltonian $H(t)$, and then the wave function's evolution follows from solving the Schrödinger equation $i\frac{d}{dt}|\Psi(t)\rangle = H(t)|\Psi(t)\rangle$[30]. This transformation not only provides a practical way to dynamically modulate the coupling strength but also enables the extraction of the Zak phase via time-dependent control.

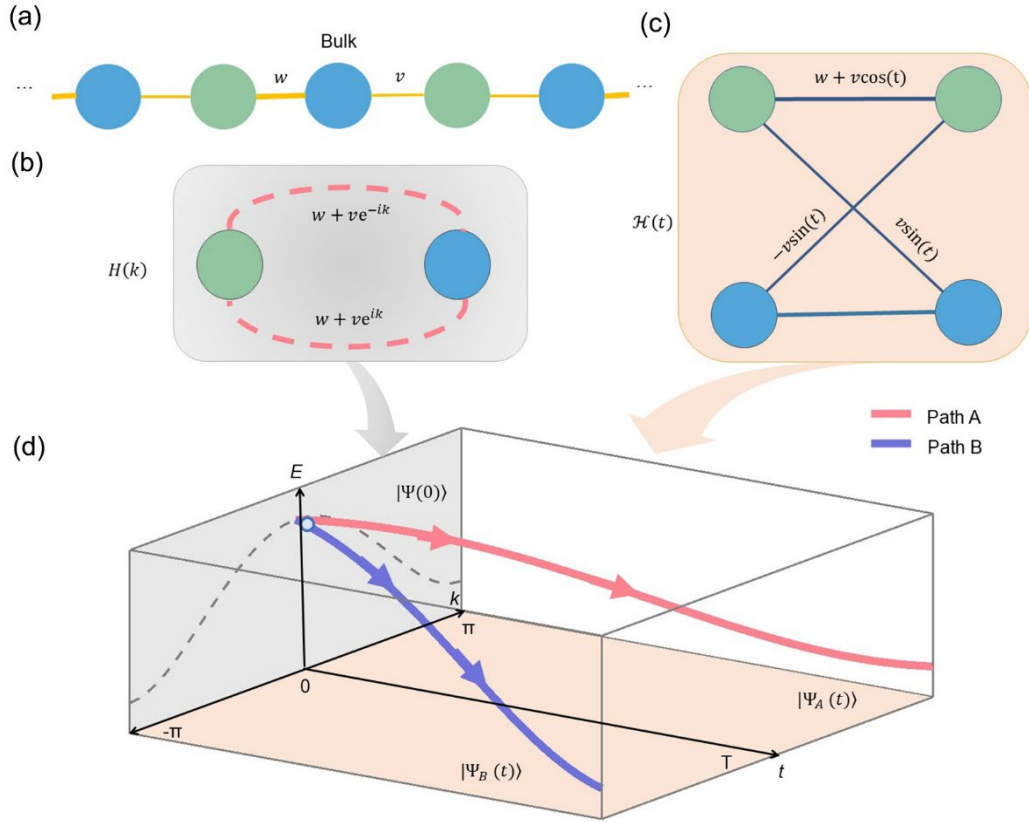

Fig.1(a) The standard SSH lattice model with alternating coupling strengths $w$ and $v$. (b) Momentum-space representation of the SSH Hamiltonian $H(k)$, showing the coupling terms as $w + ve^{\pm ik}$. (c) Corresponding time-dependent matrix $\mathcal{H}(t)$ implemented experimentally, where coupling amplitudes are dynamically modulated as functions of time. (d) Conceptual representation of the experimental procedure: the initial state $|\Psi(0)\rangle$ evolves along two symmetric trajectories (Path A and Path B) in momentum–time space. By measuring and comparing the accumulated phases along these two paths, the dynamical phases cancel out, enabling direct extraction of the geometric Zak phase.

When directly measuring the Zak phase using a time-dependent Hamiltonian $H(t)$, one typically introduces a complex component in the coupling terms, which can be difficult to implement in practical experimental setups. Consequently, additional design considerations become necessary—such as dynamic modulation in the time domain[21,31], phase-shifting interferometric network (Ramsey or Mach–Zehnder)[18,32], and helically modulated waveguide array[33,34]. Here, we propose a "dimensional extension" approach, which involves adding an extra pair of auxiliary energy levels (i.e., additional degrees of freedom) in the experiment. By replacing $\sqrt{-1}$ with the real matrix $\begin{pmatrix} 0 & 1 \\ 1 & 0 \end{pmatrix}$, the complex Hamiltonian $H(t)$ can be transformed into a higher-dimensional real matrix $\begin{bmatrix} Re(H(t)) & Im(H(t)) \\ -Im(H(t)) & Re(H(t)) \end{bmatrix}$[35-38] (see Fig. 1(c)). Upon expansion and reorganization, one obtains

$$\mathcal{H}(t) = \begin{bmatrix} 0 & 0 & d_x(t) & d_y(t) \\ 0 & 0 & -d_y(t) & d_x(t) \\ d_x(t) & -d_y(t) & 0 & 0 \\ d_y(t) & d_x(t) & 0 & 0 \end{bmatrix}, \qquad (3)$$

where $d_x(t) = w + v\cos((k(t))$, $d_y(t) = v\sin((k(t))$. We further derive the relationship between the eigenstates of $H(t)$ and $\mathcal{H}(t)$ (see Methods)

$$|\mu(t)\rangle = \begin{pmatrix} a(t) + ib(t) \\ c(t) + id(t) \end{pmatrix} \Leftrightarrow |\mathfrak{u}(t)\rangle = \begin{pmatrix} a(t) \\ b(t) \\ c(t) \\ d(t) \end{pmatrix}. \tag{4}$$

Therefore, in a four-level system, by dynamically modulating the real coupling strengths $(d_x(t), \pm d_y(t))$, the Zak phase accumulated during the adiabatic evolution process of the SSH model can be accurately simulated and measured. Under adiabatic evolution from the $n$th eigenstate $|\Psi_n(0)\rangle$, the state evolves as

$$|\Psi_n(t)\rangle = e^{i\phi_g(t)} e^{-i\int E(k(t))dt} |\Psi_n(0)\rangle, \tag{5}$$

where $-\int E(k(t))dt$ represents the dynamical phase $\phi_d(t)$ and $\phi_g(t)$ represents the geometric (Berry) phase. Drawing inspiration from the Aharonov–Bohm (AB) effect[6,39,40], we extract the geometric phase from the relative phase accumulated along two symmetric evolution paths(see Fig. 1(d)). This symmetry ensures the cancellation of dynamical phases while preserving the geometric phase due to the topological nature of the wavefunction. Thus, the phase difference $\Delta\phi_n(t) = \phi_{g,A}(t) - \phi_{g,B}(t)$ is inherently independent to the global phase of the initial state. This strategy effectively eliminates dynamical contributions in both two-level and more complex four-level systems (see Supplementary Material), enabling a more robust and precise extraction of the geometric phase while mitigating external disturbances.

**Experimental Observation of the Zak Phase**

We construct an electroacoustic experimental setup comprising four metal cavities, each with a first-order resonant frequency of $f_0 = 1955$ Hz, to serve as the lattice units (see Fig. 2). By employing a gain circuit that reduces the decay rate, the quality factor of the acoustic system is significantly improved, allowing the acoustic state to maintain coherence for a longer duration during evolution. VCA enables dynamic modulation of $\kappa$, thereby implementing the time-dependent evolution prescribed by the theoretical model. Leveraging the precise control described above, we systematically vary the evolution parameters to validate the system's adiabatic behavior. In our experimental system, adiabaticity is governed chiefly by two parameters—the total evolution time $T$ and the energy band gap. Numerical simulations show that increasing either the evolution time $T$ or the band gap improves the system's adiabaticity (see Fig. 3(a)). However, in the acoustic system, the high viscosity of the air medium and thermal conduction losses causes rapid energy decay, while scattering and radiation losses in the structure further reduce the quality factor. Although the self-gain feedback circuit has increased the $Q$ value by several orders of magnitude, effectively compensating for energy losses, the intrinsic losses in the medium and structural limitations still hinder further improvement of the $Q$ value. Taking these factors into account, we optimize the evolution time to $T = 0.5\,s$, striking a balance between effective geometric phase accumulation and stable signal quality. In our experiment, we prepare the initial state $[1\ \ 1\ \ 1\ \ 1]^T$ by applying an excitation at $f_0$, and measure the accumulated phase under various $(w, v)$ parameter combinations.

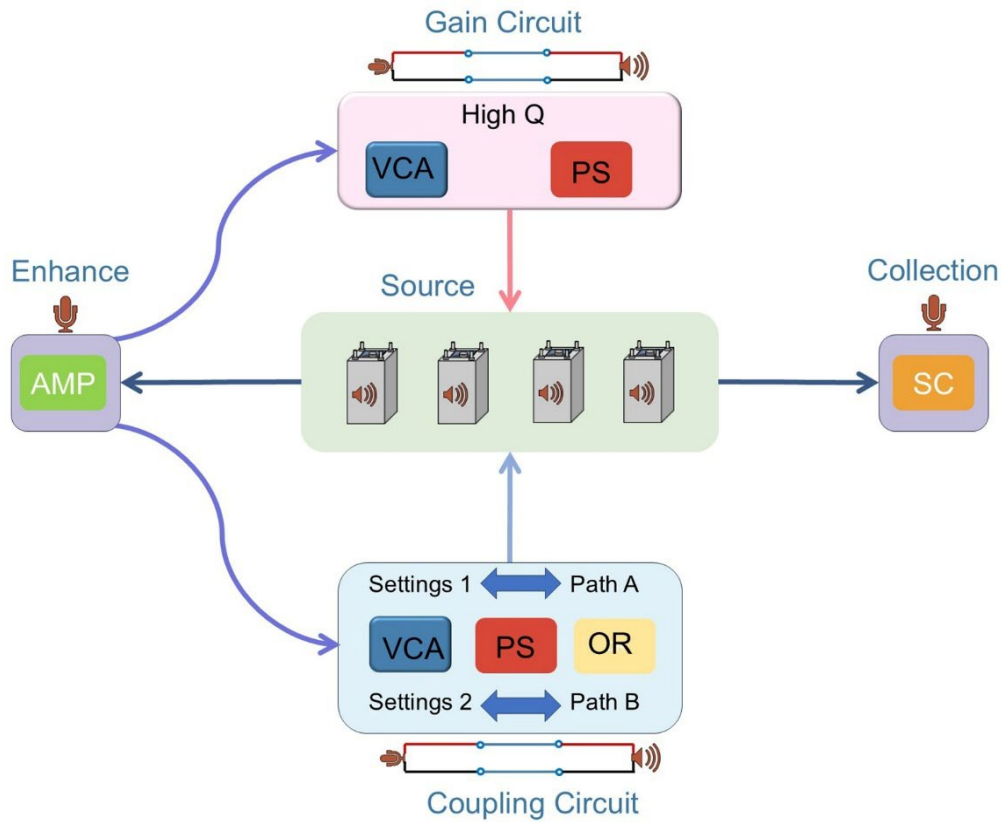

Fig.2 Electromechanical cavity experimental system. The signal within the acoustic cavity is first amplified by a primary amplifier (AMP) and then routed into both a gain

loop, which enhances the $Q$ factor, and a coupling circuit. The gain loop, consisting of a programmable voltage-controlled amplifier (VCA) and a phase shifter (PS), is used to construct a high-quality acoustic cavity. The coupling circuit simulates different evolution paths (Path A and Path B) via various configurations of circuit elements—including the VCA, PS, and an optocoupler relay (OR)—under Settings 1 and 2. Finally, a collection module uses a sound card (SC) to acquire and analyze the signal for comprehensive observation and analysis of the Zak phase.

The experimental results demonstrate that the geometric phase difference remains robust across various parameter sets, even in the presence of experimental imperfections such as noise or drift (see Fig. 3(b)). For a more detailed analysis, two extreme cases— $(w, v) = (1, 5)$ and $(5, 1)$ —are chosen, with their respective voltage and coupling profiles (see Figs. 3(c) and 3(d). Where the system is in a topologically trivial regime, the geometric phase remains close to zero, with no noticeable separation between the two paths (see Fig. 3(c)). This is also evident in the time-domain output signals, where the final signals remain in phase alignment. Conversely, in the nontrivial phase, a clear difference of $\pi$ becomes evident, driven by the underlying topological invariant (see Fig. 3(d)). This comparison not only highlights the role of topological invariants in governing waveform evolution but also provides intuitive experimental evidence of topological phase transitions. Despite unavoidable losses and coupling imperfections, the combination of self-gain feedback and voltage-controlled modulation extended the acoustic coherence time sufficiently to allow accurate geometric-phase measurements within the $0.5s$ time window. These results verify the feasibility of using electromechanical cavity systems to study topological geometric phases and lay a solid foundation for future exploration of more complex coupling networks.

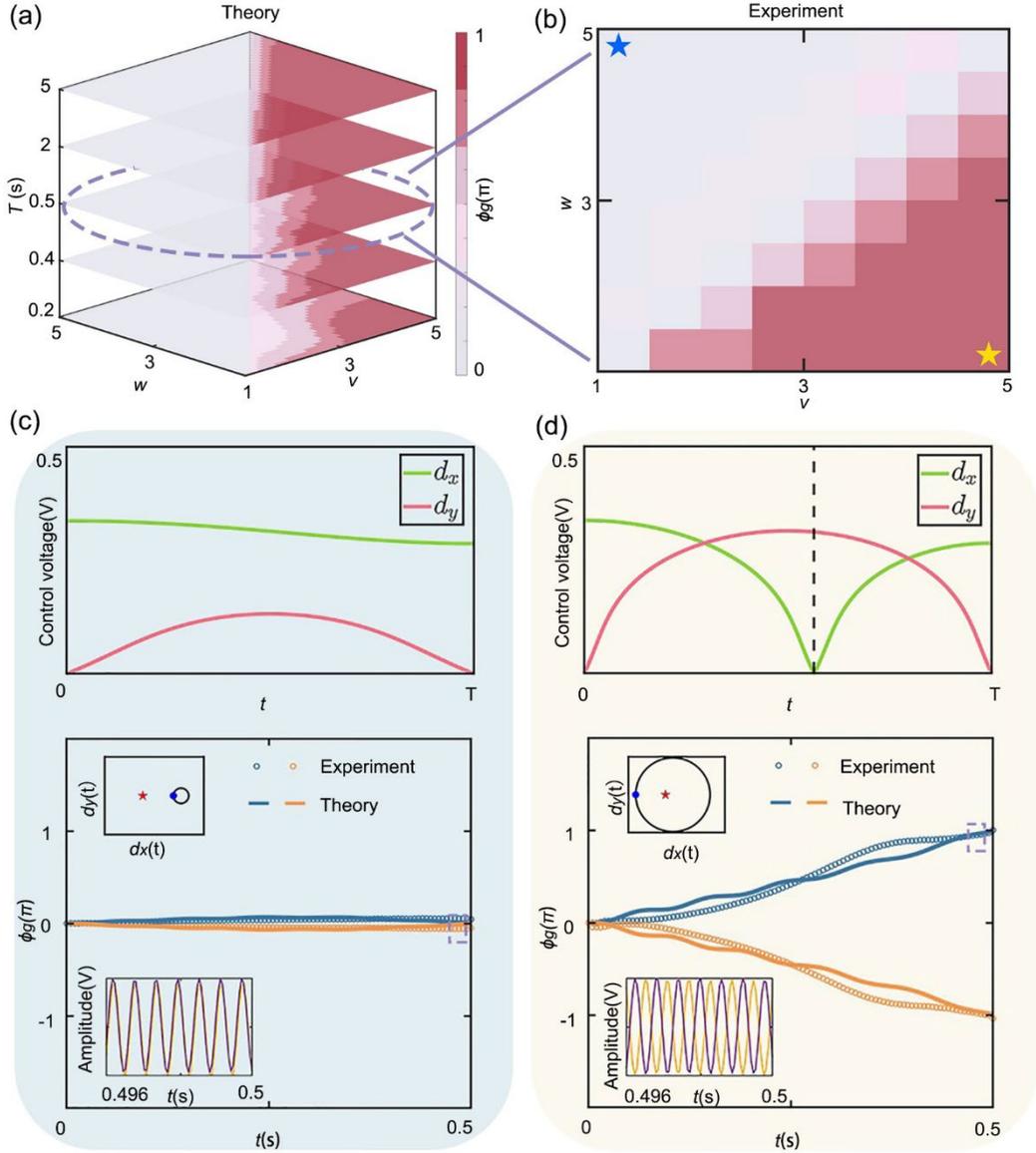

Fig. 3. Experimental characterization of geometric phase accumulation and topological phase transitions. (a) Theoretical analysis of geometric phase accumulation $\phi_g(t)$ as a function of coupling parameters $(w, v)$ and evolution time $T$. The color intensity indicates the magnitude of the geometric phase (normalized by $\pi$). (b) Corresponding experimental verification of the phase accumulation in parameter space $(w, v)$, demonstrating good agreement with theoretical predictions. Green stars represent parameter combinations selected for detailed investigation. The top of (c), (d) Experimentally implemented control voltages for coupling parameters $d_x(t)$ and $d_y(t)$, corresponding to trivial $(W = 0)$ and nontrivial $(W = 1)$ topological phases, respectively, and the dashed lines indicating the switch between positive and negative coupling. The bottom of (c), (d) Experimental (circles) and theoretical (lines) results of geometric phase accumulation for the trivial (c) and nontrivial (d) cases. Insets show trajectories of coupling parameters in the complex plane (top left) and zoom-in time-domain signal comparisons between theory and experiment (bottom). A negligible phase accumulation is observed in the trivial phase (c), while a clear linear accumulation of geometric phase approaching $\pi$ occurs in the nontrivial case (d).

To further illustrate the topological nature of the Zak phase in the SSH model, it is instructive to consider its equivalent representation in terms of the winding number[40,41]. In the momentum-space Hamiltonian, the off-diagonal component can be written as $q(k) = d_x(k) + id_y(k)$, which varies continuously with $k$ in $[-\pi, \pi]$. This defines a closed curve traced out by $q(k)$ in the complex plane. The argument around this curve is known as the winding number, given by $W = \frac{1}{2\pi} \int_{-\pi}^{\pi} \frac{d}{dk} \arg[q(k)] \, dk$. It serves as a topological invariant that is equivalent to the Zak phase up to a factor of $\pi$, and provides a geometric visualization of the system's topology. When $W = 0$, the loop does not enclose the origin, indicating a topologically trivial phase (see the inset of Fig. 3(c)). In contrast, when $W = 1$, the closed loop encloses the origin, marking a topologically nontrivial phase with a quantized Zak phase of $\pi$ (see the inset of Fig. 3(d)).

**Zak Phase Measurement of the SSH Model with Next-Nearest-Neighbor Coupling**

While the standard SSH model supports only a winding number of 0 or 1, the concept of the winding number naturally extends to values greater than 1. To investigate how higher winding numbers affect the Zak phase—and to experimentally probe the corresponding geometric phase accumulation—we extend the Hamiltonian by introducing a next-nearest-neighbor coupling term $J$[23,42,43] (see Fig. 4(a))

$$\widetilde{H}(k) = \begin{bmatrix} 0 & w + ve^{-ik} + Je^{-i2k} \\ w + ve^{ik} + Je^{i2k} & 0 \end{bmatrix}. \tag{6}$$

The $\widetilde{H}(k)$ contains higher-order terms with $e^{\pm i2k}$ factors, nevertheless, a time-dependent real-valued matrix can still be constructed by arranging the real and imaginary parts (see Fig. 4(b))

$$\widetilde{\mathcal{H}}(t) = \begin{bmatrix} 0 & 0 & d_{x1}(t) & d_{y1}(t) \\ 0 & 0 & -d_{y1}(t) & d_{x1}(t) \\ d_{x1}(t) & -d_{y1}(t) & 0 & 0 \\ d_{y1}(t) & d_{x1}(t) & 0 & 0 \end{bmatrix}, \quad (7)$$

where $d_{x1}(t) = w + v\cos((k(t))) + J\cos(2(k(t)))$, $d_{y1}(t) = v\sin((k(t))) + J\sin(2(k(t)))$. In this extended model, the inclusion of a next-nearest-neighbor coupling term $J$ introduces exponential factors of $e^{\pm i2k}$ into the Hamiltonian. These terms effectively double the rate of phase variation compared to the nearest-neighbor contributions, leading to a more complex momentum dependence on the local phase $\tilde{\theta}(k) = \arg(w + ve^{ik} + Je^{i2k})$. Critically, varying $J$ enriches the system's topological landscape by reshaping the phase accumulation trajectory and enabling multiple transitions. Thus, this tunability accommodates both conventional ($W = 0, 1$) and higher ($W = 2$) winding number configurations (see Fig. 4(c)).

In the experiment, following the previously proposed measurement scheme, we investigated the influence of different winding numbers on the geometric phase (see Fig. 4(d)). After introducing the next-nearest-neighbor coupling term $J$, the effective coupling relations $(d_{x1}(t), \pm d_{y1}(t))$ become more complicated. By carefully tuning the parameters, we map out states with higher winding numbers in hardware, enabling precise modulation of time-dependent coupling strengths (see Figs. 4(e–g)). For the trivial case with $W = 0$ and coupling parameters $(4,1,1)$, the dynamics are dominated by the nearest-neighbor interactions. The measured geometric phase $\phi_h(t)$ remains near zero throughout the evolution, showing no significant accumulation, consistent with a trajectory that does not encircle the topological singularity (see Fig. 4(e)). When the parameters are set to $(1,4,1)$, the system enters a nontrivial regime with $W = 1$. Here, $\phi_h(t)$ grows linearly in time and reaches approximately $\pi$ at

$t \approx 0.5s$, indicating that the parameter winding is once around the topological singularity (see Fig. 4(f)). Finally, further increasing the long-range coupling to $(1,1,4)$ drives the system into a non-trivial topological phase characterized by a higher winding number. In this regime, $\phi_h(t)$ exhibits a rapid global accumulation, ultimately approaching $2\pi$, signifying a $W = 2$ winding (see Fig. 4(g)). These results reveal a clear relationship between the accumulated geometric phase and the winding number. More importantly, they demonstrate that long-range coupling not only expands the accessible topological parameter space but also enables the realization of richer topological structures and higher winding $W$ lobes.

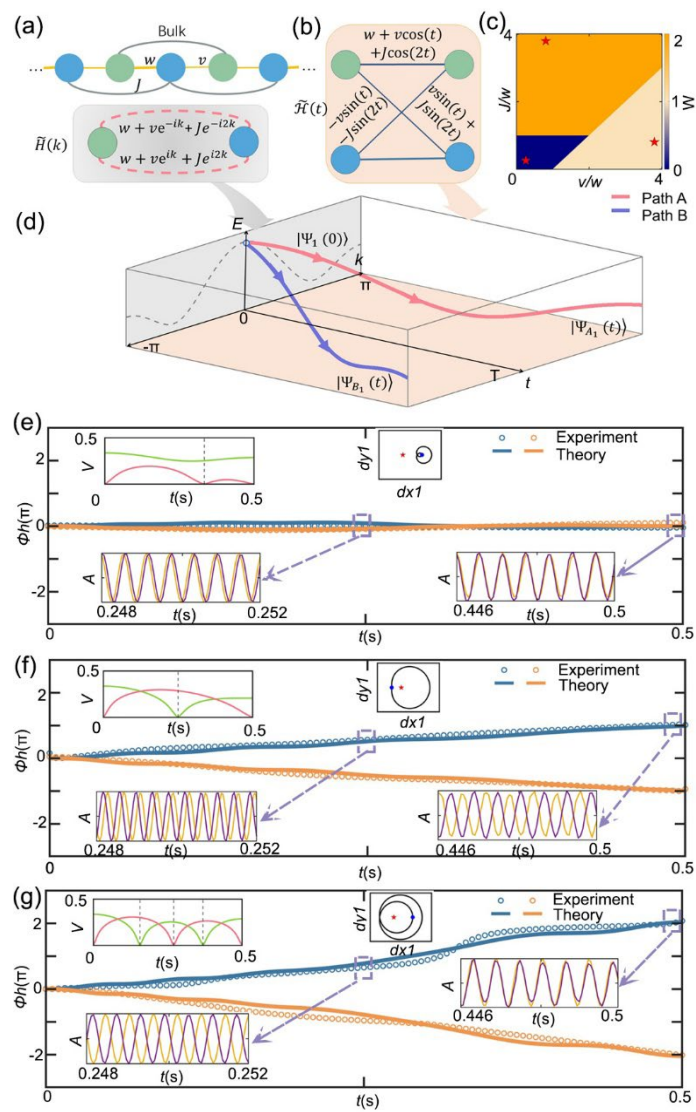

Fig. 4 Experimental demonstration of higher winding numbers and geometric phase measurement. (a) Extended SSH model with next-nearest-neighbor coupling $J$, and its momentum-space Hamiltonian representation, highlighting the additional coupling terms $Je^{\pm i2k}$. (b) Corresponding experimental implementation of the time-dependent matrix with dynamically modulated couplings. (c) The phase diagram in parameter space $(v/w, J/w)$, illustrating regions with different winding numbers $W = 0, 1, 2$. The red stars indicate experimentally explored parameter points. (d) Conceptual diagram of the experimental scheme: initial states evolve along two symmetric paths (Path A and Path B) in momentum–time space to extract the Zak phase for higher winding cases. (e)–(g) Experimental results of geometric phase accumulation $\phi_h(t)$ for different winding-number cases: (e) Trivial case ($W = 0$, parameters: $w:v:J = 4:1:1$), minimal phase accumulation, no winding around topological singularity; (f) Nontrivial SSH-like case ($W = 1$, parameters: $w:v:J = 1:4:1$), geometric phase accumulates linearly up to $\pi$; (g) Higher winding case ($W = 2$, parameters: $w:v:J = 1:1:4$), geometric phase rapidly accumulates, reaching approximately $2\pi$. Insets within (e)–(g) panels show experimental control voltages (top left), where the dashed line indicates the switching between positive and negative coupling, the red line represents $d_{x1}$ and the green line corresponds to $d_{y1}$, and the trajectory of coupling parameters in the complex plane (top right). Zoom-in panels (bottom) further confirm the excellent agreement between experimental data and theoretical predictions.

**Discussion**

In summary, we have experimentally demonstrated a direct measurement scheme for the Zak phase based on an interference approach, implemented in an electromechanical cavity system. By engineering symmetric evolution paths in parameter space and eliminating dynamical phase contributions through Aharonov–Bohm–inspired phase cancellation, we successfully extracted the geometric phase without the need for a stable reference frame. The combination of voltage-controlled coupling modulation and self-gain feedback circuits significantly enhanced the coherence time, enabling robust tracking of geometric phase accumulation throughout the evolution. In addition, our platform not only realizes the conventional SSH topological phases with winding numbers 0 or 1, but also enables the observation of topological invariants with $W = 2$, achieved by introducing next-nearest-neighbor coupling. The measured geometric phase exhibits clear linear scaling with the winding number, providing compelling evidence for the correspondence between bulk topological invariants and geometric phase accumulation. These results confirm the feasibility, precision, and extensibility of our system for investigating topological properties in synthetic dimensions. The demonstrated method provides a generic framework for probing geometric and topological properties in other classical wave systems, such as acoustic metamaterials, mechanical lattices, and circuit-based simulators. Moreover, it provides a versatile testbed for designing and investigating more sophisticated topological models, encompassing higher-dimensional systems, non-Hermitian effects, and dynamically driven topological phases.

**Methods**

**Experimental Setup**

In the experimental platform's circuit setup, the coupling strength can be dynamically adjusted based on experimental requirements through programmable control of the VCA using a controllable DC voltage. To compensate for the phase offset introduced by the primary amplifier board and VCA in signal transmission, as well as the coupling reversal effect, the system incorporates a phase shifter, effectively avoiding experimental errors caused by phase mismatches. Furthermore, the phase shifter can also adjust different evolution paths in real-time, ensuring the stability and reliability of the system. To further improve the response speed of coupling switches and transitions, the system employs a control circuit with optocoupler relays. Through this series of carefully designed components, the experimental platform is capable of accurately simulating system evolution under various dynamic conditions, thereby providing a solid foundation for experimental research.

The sound source signal is generated by a signal generator (RIGOL DG2052) and then transmitted to a speaker (balanced armature drivers, Knowles Corporation 30120) to produce a stable sound wave signal. Then, a high-sensitivity microphone (microphone head) is employed to collect the sound signal. The microphone, connected to a sound card, converts the collected sound pressure signal into a digital signal in real-time and transmits it to the computing system for processing.

**Comparative Analysis of Models Preceding and Following Dimensional Extension**

For the SSH model, the Pauli-matrix formulation is given by $H(t) = d_x(t)\sigma_x + d_y(t)\sigma_y$, where $d_x(t) = w + v\cos(t)$, $d_y(k) = v\sin(t)$, $\sigma_x = \begin{pmatrix} 0 & 1 \\ 1 & 0 \end{pmatrix}$, $\sigma_y = \begin{pmatrix} 0 & -i \\ i & 0 \end{pmatrix}$. The eigenstate of $H(t)$ is given by $|\mu(t)\rangle = \begin{pmatrix} u_1(t) \\ u_2(t) \end{pmatrix}$, which can be decomposed as

$$u_1(t) = a(t) + ib(t), \quad u_2(t) = c(t) + id(t). \tag{8}$$

Then, the original complex eigenvalue equation $H(t)|\mu(t)\rangle = E(t)|\mu(t)\rangle$. Following Eq. (10), the eigenvalue equation can be written as

$$(d_x(t)\sigma_x + d_y(t)\sigma_y) \begin{pmatrix} a(t) + ib(t) \\ c(t) + id(t) \end{pmatrix} = E(t) \begin{pmatrix} a(t) + ib(t) \\ c(t) + id(t) \end{pmatrix}. \tag{9}$$

By separating this equation into its real and imaginary parts, we obtain two independent sets of real equations:

$$d_x(t)\sigma_x \begin{pmatrix} a(t) \\ c(t) \end{pmatrix} + d_y(t)\sigma_{y'} \begin{pmatrix} b(t) \\ d(t) \end{pmatrix} = E(t) \begin{pmatrix} a(t) \\ c(t) \end{pmatrix},$$

$$d_x(t)\sigma_x \begin{pmatrix} ib(t) \\ id(t) \end{pmatrix} - d_y(t)i\sigma_{y'} \begin{pmatrix} a(t) \\ c(t) \end{pmatrix} = E(t) \begin{pmatrix} ib(t) \\ id(t) \end{pmatrix}, \tag{10}$$

where $\sigma_{y'} = i\sigma_y = \begin{pmatrix} 0 & 1 \\ -1 & 0 \end{pmatrix}$. Rearranging these equations, we can write them in the following block form:

$$\begin{bmatrix} d_x(t)\sigma_x & d_y(t)\sigma_{y'} \\ -d_y(t)\sigma_{y'} & d_x(t)\sigma_x \end{bmatrix} \begin{pmatrix} a(t) \\ c(t) \\ b(t) \\ d(t) \end{pmatrix} = E(t) \begin{pmatrix} a(t) \\ c(t) \\ b(t) \\ d(t) \end{pmatrix}. \quad (11)$$

Thus, we have the correspondence between the eigenvalue equations of the original and the mapped systems:

$$H(t)|\mu(t)\rangle = E(t)|\mu(t)\rangle \Leftrightarrow \rho(H(t))|\tilde{\mu}(t)\rangle = E(t)|\tilde{\mu}(t)\rangle,$$

$$|\mu(t)\rangle = \begin{pmatrix} u_1(t) \\ u_2(t) \end{pmatrix} \Leftrightarrow |\tilde{\mu}(t)\rangle = \begin{pmatrix} a(t) \\ c(t) \\ b(t) \\ d(t) \end{pmatrix}. \quad (12)$$

Here, the real matrix is defined as

$$\mathbb{H}(t) = \begin{bmatrix} d_x(t)\sigma_x & d_y(t)\sigma_{y'} \\ -d_y(t)\sigma_{y'} & d_x(t)\sigma_x \end{bmatrix} = \begin{bmatrix} 0 & d_x(t) & 0 & d_y(t) \\ d_x(t) & 0 & -d_y(t) & 0 \\ 0 & -d_y(t) & 0 & d_x(t) \\ d_y(t) & 0 & d_x(t) & 0 \end{bmatrix}. \quad (13)$$

The original complex eigenstate is split into two real subspaces, which leads to a doubling of the eigenvalue multiplicity. In practice, each pair of real eigenvalues strictly corresponds to the same complex eigenvalue $E(t)$ of the original Hamiltonian, thereby preserving the band structure of the SSH model. In our numerical simulations and analysis, we adopt an alternative real matrix form obtained after an appropriate basis transformation. This alternative representation, denoted by $\mathcal{H}(t)$. It can be verified that the relationship between $\mathcal{H}(t)$ and the earlier matrix $\mathbb{H}(t)$ is achieved through a basis transformation $P$, such that

$$(P\mathbb{H}(t)P^{-1})P \begin{pmatrix} a(t) \\ c(t) \\ b(t) \\ d(t) \end{pmatrix} = E(t)P \begin{pmatrix} a(t) \\ c(t) \\ b(t) \\ d(t) \end{pmatrix}, P = \begin{bmatrix} 1 & 0 & 0 & 0 \\ 0 & 0 & 1 & 0 \\ 0 & 1 & 0 & 0 \\ 0 & 0 & 0 & 1 \end{bmatrix}. \quad (14)$$

After the basis transformation, the real eigenstate of $\mathcal{H}(t)$ is given by

$$|\mathfrak{u}(t)\rangle = P|\tilde{\mathfrak{u}}(t)\rangle = \begin{pmatrix} a(t) \\ b(t) \\ c(t) \\ d(t) \end{pmatrix}. \tag{15}$$

The difference between the two real-matrix representations is solely due to the choice of basis; the physical content remains unchanged. Therefore, the transformation not only represents a mathematical rearrangement but also crucially preserves the band characteristics of the original SSH model.

In the complex Hamiltonian, the expression $d_x(t) + id_y(t) = r(t)e^{i\theta(t)}$ simultaneously encodes the amplitude $r(t)$ and the phase $\theta(t)$. When constructing the real matrix $\mathcal{H}(t)$ via the mapping, the coupling terms appear in the form of $\left(d_x(t), \pm d_y(t)\right)$. Owing to the isomorphism, the real matrix effectively 'reconstructs' the two components of the original complex number, so that when the mapped real eigenstate is reassembled into a complex form, one inevitably recovers the same complex quantity $\left(d_x(t), \pm d_y(t)\right)$, thereby fully preserving the accumulation of the phase information $e^{i\theta(t)}$ during the evolution in the Brillouin zone. This correspondence between theory and numerical results ensures that the phase evolution and band topology in the SSH model are implemented and validated in a more intuitive and rigorous manner across a broader range of systems.

**Data availability**

The main data supporting the findings of this study are available within this letter and its supplementary information.

**Code availability**

The code used to analyze the data and generate the plots for this paper is available from the corresponding author upon request.

**References**

1. Hasan M Z, Kane C L. Colloquium: Topological insulators. *Rev. Mod. Phys*. **82**, 3045 (2010).
2. Qi X L, Zhang S C. Topological insulators and superconductors. *Rev. Mod. Phys.* **83**, 1057 (2011).
3. Lu L, Joannopoulos J D, Soljačić M. Topological photonics. *Nat. Photonics* **8**, 821–829 (2014).
4. Ma G, Xiao M, Chan C T. Topological phases in acoustic and mechanical systems. *Nat. Rev. Phys*. **1**, 281–294 (2019).
5. Thouless D J, Kohmoto M, Nightingale M P & den Nijs M. Quantized Hall Conductance in a Two-Dimensional Periodic Potential. *Phys. Rev. Lett*. **49**, 405 (1982).
6. Berry M V. Quantal phase factors accompanying adiabatic changes. *Proc. R. Soc. A* **392**, 45 (1984).
7. Zak J. Berry's phase for energy-bands in solids. *Phys. Rev. Lett*. **62**, 2747 (1989).
8. Hatsugai Y. Chern number and edge states in the integer quantum Hall effect. *Phys. Rev. Lett*. **71**, 3697 (1993).
9. Kane C L, Mele E J. $Z_2$ Topological Order and the Quantum Spin Hall Effect. *Phys. Rev. Lett*. **95**, 146802 (2005).
10. Sompet P. et al. Realizing the symmetry-protected Haldane phase in Fermi–Hubbard ladders. *Nature* **606**, 484 (2022).
11. Redon Q. et al. Realizing the entanglement Hamiltonian of a topological quantum Hall system. *Nat. Commun.* **15**, 10086 (2024).
12. Chang C. et al. Coupling-Controlled Photonic Topological Ring Array. *ACS Photonics* **11**, 5260 (2024).
13. Huang L. et al. Hyperbolic photonic topological insulators. *Nat. Commun.* **15**, 1647 (2024).
14. Meng Y. et al. Spinful Topological Phases in Acoustic Crystals with Projective PT Symmetry. *Phys. Rev. Lett.* **130**, 026101 (2023).
15. Sun X C. et al. Ideal acoustic quantum spin Hall phase in a multi-topology platform. *Nat. Commun.* **14**, 952 (2023).

16. Zhang W. et al. Experimental Observation of Higher-Order Topological Anderson Insulators. *Phys. Rev. Lett.* **126**, 146802 (2021).

17. Liu S. et al. Non-Hermitian Skin Effect in a Non-Hermitian Electrical Circuit. *Research* **2021**, 5608038 (2021).

18. Atala M. et al. Direct measurement of the Zak phase in topological Bloch bands. *Nat. Phys.* **9**, 795 (2013).

19. Bharath H M, Boguslawski M, Barrios M, et al. Exploring Non-Abelian Geometric Phases in Spin-1 Ultracold Atoms. *Phys. Rev. Lett.* **123**, 173202 (2019).

20. Singhal Y. et al. Measuring the adiabatic non-Hermitian Berry phase in feedback-coupled oscillators. *Phys. Rev. Res.* **5**, L032026 (2023).

21. Chen Z X. et al. Direct Measurement of Topological Invariants through Temporal Adiabatic Evolution of Bulk States in the Synthetic Brillouin Zone. *Phys. Rev. Lett.* **134**, 136601 (2025).

22. Jiao Z Q. et al. Experimentally Detecting Quantized Zak Phases without Chiral Symmetry in Photonic Lattices. *Phys. Rev. Lett.* **127**, 147401 (2021).

23. Hsu H C, Chen T W. Topological Anderson insulating phases in the long-range Su-Schrieffer-Heeger model. *Phys. Rev. B* **102**, 205425 (2020).

24. Liu H. et al. Acoustic spin-Chern topological Anderson insulators. *Phys. Rev. B* **108**, L161410 (2023).

25. Peng T. et al. Structural disorder–induced topological phase transitions in quasicrystals. *Phys. Rev. B* **109**, 195301 (2024).

26. Li Y. et al. Interaction-induced breakdown of chiral dynamics in the Su-Schrieffer-Heeger model. *Phys. Rev. Res.* **5**, L032035 (2023).

27. Chen Z X. et al. Transient logic operations in acoustics through dynamic modulation. *Phys. Rev. Appl.* **21**, L011001 (2024).

28. Chen Z X. et al. Robust temporal adiabatic passage with perfect frequency conversion between detuned acoustic cavities. *Nat. Commun.* **15**, 1478 (2024).

29. Asbóth J K, Oroszlány L, Pályi A. A Short Course on Topological Insulators: Band-Structure Topology and Edge States in One and Two Dimensions. Cham: Springer, 2016.

30. Chen Z X. et al. Emulation of Schrödinger dynamics with metamaterials. *Sci. Bull*. **70**, 8 (2025).

31. Zhang Q C. et al. Observation of Acoustic Non-Hermitian Bloch Braids and Associated Topological Phase Transitions. *Phys. Rev. Lett*. **130**, 017201 (2023).

32. On M B. et al. Programmable integrated photonics for topological Hamiltonians. *Nat. Commun.* **15**, 629 (2024).

33. Rechtsman M C. et al. Photonic Floquet topological insulators. Nature **496**, 196–200 (2013).

34. Parto M. et al. Observation of twist-induced geometric phases and inhibition of optical tunneling via Aharonov–Bohm effects. *Sci. Adv*. **5**, eaau8135 (2019).

35. Barlas Y, Prodan E. Topological classification table implemented with classical passive metamaterials. *Phys. Rev. B* **98**, 094310 (2018).

36. Shao L B. et al. Spinless mirror Chern insulator from projective symmetry algebra. *Phys. Rev. B* **108**, 205126 (2023).

37. Wu S Q. et al. Observation of D-class topology in an acoustic metamaterial. *Sci. Bull.* **69**, 893 (2024).

38. Sun X C, Wang J B, He C & Chen Y F. Non-Abelian Topological Phases and Their Quotient Relations in Acoustic Systems. *Phys. Rev. Lett.* **132**, 216602 (2024).

39. Aharonov Y, Bohm D. Significance of electromagnetic potentials in the quantum theory. *Phys. Rev.* **115**, 485 (1959).

40. Simon B. Holonomy, the quantum adiabatic theorem, and Berry phase. *Phys. Rev. Lett.* **51**, 2167(1983).

41. Schnyder A P, Ryu S, Furusaki A & Ludwig A W W. Classification of topological insulators and superconductors in three spatial dimensions. *Phys. Rev. B* **78**, 195125 (2008).

42. Maffei M. et al. Topological characterization of chiral models through their long-time dynamics. *New J. Phys*. **20**, 013023 (2018).

43. Rufo S, Lopes N, Continentino M A & Griffith M A R. Multicritical behavior in topological phase transitions. *Phys. Rev. B* **100**, 195432 (2019).

**Acknowledgements**

This work is supported by the National Key Research and Development Program of China (Grant No.2023YFA1406900) and the National Natural Science Foundation of China (Nos. 1247043673 and 12404506).

**Author contributions**

Z.-G.C. conceived the idea. Z.-G.C. and G.-C.H. developed the theory. G.-C.H. and Z.-X.C. performed the experiment. All authors contributed to analyzing the data and writing the manuscript.

**Competing interests**

The authors declare no competing interests.

**Additional information**

Correspondence and requests for materials should be addressed to Zeguo Chen.